\documentclass[conference,a4paper]{IEEEtran}
\IEEEoverridecommandlockouts
\usepackage[final]{graphicx}
\usepackage{times}
\usepackage{amsfonts}
\usepackage{cite}
\usepackage{epsfig}
\usepackage{amssymb}
\usepackage{amsmath}
\usepackage{color}
\usepackage{cite}
\usepackage{amsbsy}
\usepackage{verbatim}

\newcommand{\beq}{\begin{equation}}
\newcommand{\eeq}{\end{equation}}
\newcommand{\ben}{\begin{enumerate}}
\newcommand{\een}{\end{enumerate}}
\newcommand{\bit}{\begin{itemize}}
\newcommand{\eit}{\end{itemize}}

\newtheorem{example}{Example}

\begin{document}
\setlength\unitlength{1mm}

\newcommand{\insertfig}[3]{
\begin{figure}[tbp]\begin{center}\begin{picture}(120,90)
\put(0,-5){\includegraphics[width=12cm,height=9cm,clip=]{#1.eps}}\end{picture}\end{center}
\caption{#2}\label{#3}\end{figure}}

\newcommand{\inserttwofig}[4]{
\begin{figure}\begin{center}\begin{picture}(150,170)
\put(10,80){\includegraphics[with=8.5cm]{#1.ps}}
\put(10, 0){\includegraphics[with=8.5cm]{#2.ps}}
\end{picture}\end{center}
\caption{#3}\label{#4}\end{figure}}


\newfont{\bma}{msbm10 scaled 800}
\newcommand{\CCm}{\mbox{\bma C}}
\newcommand{\RRm}{\mbox{\bma R}}
\newcommand{\ZZm}{\mbox{\bma Z}}
\newcommand{\FFm}{\mbox{\bma F}}
\newcommand{\GGm}{\mbox{\bma G}}
\newcommand{\EEm}{\mbox{\bma E}}

\newfont{\bb}{msbm10 scaled 1100}
\newcommand{\CC}{\mbox{\bb C}}
\newcommand{\RR}{\mbox{\bb R}}
\newcommand{\ZZ}{\mbox{\bb Z}}
\newcommand{\FF}{\mbox{\bb F}}
\newcommand{\GG}{\mbox{\bb G}}
\newcommand{\EE}{\mbox{\bb E}}

\newcommand{\GaussBin}{G}

\newfont{\bs}{msbm10 scaled 2200} 
\newcommand{\Cs}{\mbox{\bs C}} 
\newcommand{\Rs}{\mbox{\bs R}} 
\newcommand{\Zs}{\mbox{\bs Z}} 
\newcommand{\Fs}{\mbox{\bs F}} 
\newcommand{\Gs}{\mbox{\bs G}}

\newcommand{\av}{{\bf a}}
\newcommand{\bv}{{\bf b}}
\newcommand{\cv}{{\bf c}}
\newcommand{\dv}{{\bf d}}
\newcommand{\ev}{{\bf e}}
\newcommand{\fv}{{\bf f}}
\newcommand{\gv}{{\bf g}}
\newcommand{\hv}{{\bf h}}
\newcommand{\iv}{{\bf i}}
\newcommand{\jv}{{\bf j}}
\newcommand{\kv}{{\bf k}}
\newcommand{\lv}{{\bf l}}
\newcommand{\mv}{{\bf m}}
\newcommand{\nv}{{\bf n}}
\newcommand{\ov}{{\bf o}}
\newcommand{\pv}{{\bf p}}
\newcommand{\qv}{{\bf q}}
\newcommand{\rv}{{\bf r}}
\newcommand{\sv}{{\bf s}}
\newcommand{\tv}{{\bf t}}
\newcommand{\uv}{{\bf u}}
\newcommand{\wv}{{\bf w}}
\newcommand{\vv}{{\bf v}}
\newcommand{\xv}{{\bf x}}
\newcommand{\yv}{{\bf y}}
\newcommand{\zv}{{\bf z}}
\newcommand{\zerov}{{\bf 0}}
\newcommand{\onev}{{\bf 1}}


\newcommand{\Am}{{\bf A}}
\newcommand{\Bm}{{\bf B}}
\newcommand{\Cm}{{\bf C}}
\newcommand{\Dm}{{\bf D}}
\newcommand{\Em}{{\bf E}}
\newcommand{\Fm}{{\bf F}}
\newcommand{\Gm}{{\bf G}}
\newcommand{\Hm}{{\bf H}}
\newcommand{\Id}{{\bf I}}
\newcommand{\Jm}{{\bf J}}
\newcommand{\Km}{{\bf K}}
\newcommand{\Lm}{{\bf L}}
\newcommand{\Mm}{{\bf M}}
\newcommand{\Nm}{{\bf N}}
\newcommand{\Om}{{\bf O}}
\newcommand{\Pm}{{\bf P}}
\newcommand{\Qm}{{\bf Q}}
\newcommand{\Rm}{{\bf R}}
\newcommand{\Sm}{{\bf S}}
\newcommand{\Tm}{{\bf T}}
\newcommand{\Um}{{\bf U}}
\newcommand{\Wm}{{\bf W}}
\newcommand{\Vm}{{\bf V}}
\newcommand{\Xm}{{\bf X}}
\newcommand{\Ym}{{\bf Y}}
\newcommand{\Zm}{{\bf Z}}


\newcommand{\Ac}{{\cal A}}
\newcommand{\Bc}{{\cal B}}
\newcommand{\Cc}{{\cal C}}
\newcommand{\Dc}{{\cal D}}
\newcommand{\Ec}{{\cal E}}
\newcommand{\Fc}{{\cal F}}
\newcommand{\Gc}{{\cal G}}
\newcommand{\Hc}{{\cal H}}
\newcommand{\Ic}{{\cal I}}
\newcommand{\Jc}{{\cal J}}
\newcommand{\Kc}{{\cal K}}
\newcommand{\Lc}{{\cal L}}
\newcommand{\Mc}{{\cal M}}
\newcommand{\Nc}{{\cal N}}
\newcommand{\Oc}{{\cal O}}
\newcommand{\Pc}{{\cal P}}
\newcommand{\Qc}{{\cal Q}}
\newcommand{\Rc}{{\cal R}}
\newcommand{\Sc}{{\cal S}}
\newcommand{\Tc}{{\cal T}}
\newcommand{\Uc}{{\cal U}}
\newcommand{\Wc}{{\cal W}}
\newcommand{\Vc}{{\cal V}}
\newcommand{\Xc}{{\cal X}}
\newcommand{\Yc}{{\cal Y}}
\newcommand{\Zc}{{\cal Z}}


\newcommand{\bolda}{\mbox{\boldmath$a$}}
\newcommand{\boldb}{\mbox{\boldmath$b$}}
\newcommand{\boldc}{\mbox{\boldmath$c$}}
\newcommand{\boldd}{\mbox{\boldmath$d$}}
\newcommand{\bolde}{\mbox{\boldmath$e$}}
\newcommand{\boldf}{\mbox{\boldmath$f$}}
\newcommand{\boldg}{\mbox{\boldmath$g$}}
\newcommand{\boldh}{\mbox{\boldmath$h$}}
\newcommand{\boldi}{\mbox{\boldmath$i$}}
\newcommand{\boldj}{\mbox{\boldmath$j$}}
\newcommand{\boldk}{\mbox{\boldmath$k$}}
\newcommand{\boldl}{\mbox{\boldmath$l$}}
\newcommand{\boldm}{\mbox{\boldmath$m$}}
\newcommand{\boldn}{\mbox{\boldmath$n$}}
\newcommand{\boldo}{\mbox{\boldmath$o$}}
\newcommand{\boldp}{\mbox{\boldmath$p$}}
\newcommand{\boldq}{\mbox{\boldmath$q$}}
\newcommand{\boldr}{\mbox{\boldmath$r$}}
\newcommand{\bolds}{\mbox{\boldmath$s$}}
\newcommand{\boldt}{\mbox{\boldmath$t$}}
\newcommand{\boldu}{\mbox{\boldmath$u$}}
\newcommand{\boldv}{\mbox{\boldmath$v$}}
\newcommand{\boldw}{\mbox{\boldmath$w$}}
\newcommand{\boldx}{\mbox{\boldmath$x$}}
\newcommand{\boldy}{\mbox{\boldmath$y$}}
\newcommand{\boldz}{\mbox{\boldmath$z$}}


\newcommand{\alphav}{\hbox{\boldmath$\alpha$}}
\newcommand{\betav}{\hbox{\boldmath$\beta$}}
\newcommand{\gammav}{\hbox{\boldmath$\gamma$}}
\newcommand{\deltav}{\hbox{\boldmath$\delta$}}
\newcommand{\etav}{\hbox{\boldmath$\eta$}}
\newcommand{\lambdav}{\hbox{\boldmath$\lambda$}}
\newcommand{\epsilonv}{\hbox{\boldmath$\epsilon$}}
\newcommand{\nuv}{\hbox{\boldmath$\nu$}}
\newcommand{\muv}{\hbox{\boldmath$\mu$}}
\newcommand{\zetav}{\hbox{\boldmath$\zeta$}}
\newcommand{\phiv}{\hbox{\boldmath$\phi$}}
\newcommand{\psiv}{\hbox{\boldmath$\psi$}}
\newcommand{\thetav}{\hbox{\boldmath$\theta$}}
\newcommand{\tauv}{\hbox{\boldmath$\tau$}}
\newcommand{\omegav}{\hbox{\boldmath$\omega$}}
\newcommand{\xiv}{\hbox{\boldmath$\xi$}}
\newcommand{\sigmav}{\hbox{\boldmath$\sigma$}}
\newcommand{\piv}{\hbox{\boldmath$\pi$}}

\newcommand{\Gammam}{\hbox{\boldmath$\Gamma$}}
\newcommand{\Lambdam}{\hbox{\boldmath$\Lambda$}}
\newcommand{\Deltam}{\hbox{\boldmath$\Delta$}}
\newcommand{\Sigmam}{\hbox{\boldmath$\Sigma$}}
\newcommand{\Phim}{\hbox{\boldmath$\Phi$}}
\newcommand{\Pim}{\hbox{\boldmath$\Pi$}}
\newcommand{\Psim}{\hbox{\boldmath$\Psi$}}
\newcommand{\Thetam}{\hbox{\boldmath$\Theta$}}
\newcommand{\Omegam}{\hbox{\boldmath$\Omega$}}
\newcommand{\Xim}{\hbox{\boldmath$\Xi$}}

\newcommand{\argmax}{\arg\!\max}
\newcommand{\argmin}{\arg\!\min}
\newcommand{\sinc}{{\hbox{sinc}}}
\newcommand{\diag}{{\hbox{diag}}}
\renewcommand{\det}{{\hbox{det}}}
\newcommand{\trace}{{\hbox{trace}}}
\newcommand{\sign}{{\hbox{sign}}}
\renewcommand{\arg}{{\hbox{arg}}}
\newcommand{\var}{{\hbox{var}}}
\newcommand{\SINR}{\hbox{SINR}}
\newcommand{\SNR}{\hbox{SNR}}
\newcommand{\Ei}{{\rm E}_{\rm i}}
\renewcommand{\Re}{{\rm Re}}
\renewcommand{\Im}{{\rm Im}}
\newcommand{\eqdef}{\stackrel{\Delta}{=}}
\newcommand{\<}{\left\langle}
\renewcommand{\>}{\right\rangle}
\newcommand{\approxpropto}{\stackrel{\sim}{\propto}}

\title{Nonbinary Spatially-Coupled LDPC Codes on the Binary Erasure Channel}

\author{Amina Piemontese$^{\dag}$, Alexandre Graell i Amat$^{\ddag}$, and Giulio Colavolpe$^{\dag}$\\
\normalsize $\dag$Universit\`a di Parma, Dipartimento di Ingegneria dell'Informazione, Viale G. P. Usberti 181/A, Parma, Italy\\
$\ddag$ Department of Signals and Systems, Chalmers University of Technology, Gothenburg, Sweden.}
\maketitle

\begin{abstract}

We analyze the asymptotic performance of nonbinary spatially-coupled low-density parity-check (SC-LDPC) codes built on the general linear group, when the transmission takes place over the binary erasure channel. 
We propose an efficient method to derive an upper bound to the maximum a posteriori probability (MAP) threshold for nonbinary LDPC codes, and observe that the MAP performance of regular LDPC codes improves with the alphabet size. 
We then consider nonbinary SC-LDPC codes. We show that the same threshold saturation effect experienced by binary SC-LDPC codes occurs
for the nonbinary codes, hence we conjecture that the BP threshold for large
termination length approaches the MAP threshold of the underlying regular ensemble.

\end{abstract}

\section{Introduction}
Low-density parity-check (LDPC) codes are a powerful class of codes achieving rates very close to capacity for binary memoryless symmetric (BMS) channels. 
Their excellent performance, however, usually requires long block lengths over the binary field. For short-to-moderate block lengths, nonbinary LDPC codes have been shown to outperform their binary counterparts \cite{DaMa98}. For this reason, nonbinary LDPC codes designed over Galois fields of order $2^m$ ($\textrm{GF}_2^m$), where $m$ is the number of bits per symbol, 
have received a considerable interest in the last few years. Their performance under iterative decoding was analyzed in 
\cite{RaUr05,DeFo07}.
In \cite{RaUr05} the density evolution (DE) for nonbinary 
LDPC code ensembles defined with respect to the general linear 
group over the binary field was derived for the binary erasure channel (BEC). It was shown that the messages exchanged in the belief propagation (BP) decoder can be interpreted as subspaces of the
vector space $\textrm{GF}_2^m$, which need to be enumerated. Also, it was  observed in \cite{RaUr05} that the BP threshold of some nonbinary LDPC code ensembles improves up to a certain $m$ and then worsens for increasing values of $m$. 
Upper bounds to the maximum a posteriori probability (MAP) thresholds of nonbinary LDPC code ensembles were also given, and they were conjectured to be tight. Later, in \cite{RaAn11}, the Maxwell construction of \cite{MeMoUr08}, relating 
the performance of the MAP and the BP decoder, was shown to hold for nonbinary LDPC codes over $\textrm{GF}_2^2$.

Spatially-coupled LDPC (SC-LDPC) codes
\cite{FeZi99} have received notable attention in the recent years due to their outstanding performance 
for a myriad of channels and communication problems. For the BEC, it was proved in \cite{KuRiUr11} that
the BP decoding threshold of a binary SC-LDPC code achieves the
optimal MAP threshold of the underlying LDPC block
code ensemble, a phenomenon known as \textit{threshold saturation}. This result has been recently extended to BMS channels \cite{KuMeRiUr10}, and the same phenomenon has been observed for many other channels, such as the multiple access and the relay 
channel. However, despite their excellent performance for long blocks, SC-LDPC codes perform poorly for the short-to-moderate block length regime, even worse than, e.g., irregular LDPC codes.

In this paper, we consider nonbinary SC-LDPC codes for transmission over the BEC. 
To the best of our knowledge, only the recently submitted paper~\cite{UcKaSa11} addresses the construction of nonbinary SC-LDPC codes and reports some BP thresholds for the BEC. However, no analysis on the MAP threshold nor on the threshold saturation phenomenon of nonbinary SC-LDPC codes is performed in~\cite{UcKaSa11}. 
Here, to analyze the asymptotic performance of nonbinary
SC-LDPC codes of nonbinary SC-LDPC codes we first to consider nonbinary LDPC codes and, in particular, analyze their MAP threshold. 
We give a systematic and elegant way to generate all subspaces of $\textrm{GF}_2^m$ of a certain dimension, which allows us to compute the BP extrinsic information transfer (BP EXIT) curve 
for an arbitrary $m$ and subsequently an upper bound to the MAP threshold for nonbinary LDPC codes. We show that MAP threshold bound of regular ensembles improves with $m$ and approaches the Shannon limit. 
We then analyze nonbinary SC-LDPC codes and show that, contrary to regular and irregular LDPC codes for which the BP decoding threshold worsens for high values of $m$, the BP threshold of nonbinary SC-LDPC codes 
with large termination length improves with $m$ (for the values analyzed) and tends to the Shannon limit. We also show the threshold saturation phenomenon for given $m$, and we conjecture that the BP threshold of SC-LDPC codes 
with increasing termination length saturates to the MAP threshold of the underlying regular ensemble.

\section{Nonbinary LDPC codes}

We consider transmission over a BEC with erasure probability $\varepsilon$ using nonbinary LDPC codes defined over the general linear group~\cite{RaUr05}. 
The code symbols are elements of the binary vector space $\textrm{GF}_2^m$ of dimension $m$. The code block length is $n$ symbols, 
and we transmit on the BEC the $m$-tuples representing their binary image. Therefore, we interpret the codeword as 
a binary codeword of length $nm$. We denote by $x_i$ the $i$th information bit and by $y_i$ the corresponding channel output, 
which is a random variable over $\{0,1,?\}$, where symbol $?$ denotes an erasure.
The channel outputs are collected into the vector $\yv=\{ y_1,\cdots, y_{nm}\}$. 
Also, we denote by $\yv_{\sim i}$ the vector of the channel outputs when the $i$th sample is omitted. We denote a regular nonbinary LDPC code ensemble as $\mathcal{G}(d_\texttt{v},d_\texttt{c},m)$, where $d_\texttt{v}$ is the variable-node degree and $d_\texttt{c}$ is the check-node degree. Given a code in this ensemble, we associate 
to each edge of the corresponding graph a bijective linear mapping $f:\textrm{GF}_2^m \rightarrow \textrm{GF}_2^m$, chosen 
uniformly at random. The set of mappings is the general linear group $\textrm{GL}_2^m$ over the binary field, 
which is the set of all $m\times m$ invertible matrices whose entries take values on $\{0,1\}$.
The design rate $r$ of a code in the ensemble $\mathcal{G}(d_\texttt{v},d_\texttt{c},m)$ does not depend on $m$ and can be expressed as 
$r=1-\frac{d_\texttt{v}}{d_\texttt{c}}$.

In this work, we are interested in the asymptotic average performance of the regular ensemble when \mbox{$n\rightarrow \infty$}. The asymptotic performance of LDPC codes can be analyzed in terms of the MAP and BP thresholds. 
We denote the MAP and the BP thresholds by $\varepsilon^{\textrm{MAP}}$ and $\varepsilon^{\textrm{BP}}$, respectively. In the case of 
transmission over the BEC, $\varepsilon^{\textrm{MAP}}$ is the  largest channel parameter such that the normalized conditional 
entropy converges to zero. The evaluation of $\varepsilon^{\textrm{MAP}}$ is not an easy task, but an upper bound can be obtained by computing the 
asymptotic average BP EXIT curve, which corresponds to running a BP decoder on a very large graph until the decoder has reached a fixed point. 
This can be accomplished by means of the DE method~\cite{RiUr08}.
Given a code $\texttt{G}$ in the ensemble $\mathcal{G}$, the BP EXIT curve at the $\ell$th iteration is defined as
\begin{equation}\nonumber
h_\texttt{G}^{\textrm{BP},\ell}(\varepsilon)=\frac{1}{nm}\sum_iP(\hat{x}^\ell_i=?|\yv_{\sim i})\, ,
\end{equation}
where $\hat{x}^\ell_i$ is the estimate delivered by the BP decoder at the $\ell$th iteration.
The asymptotic average BP EXIT curve of the ensemble is defined as
\begin{equation}\label{e:BP EXIT}
h^{\textrm{BP}}(\varepsilon)=\lim_{\ell\rightarrow \infty} \lim_{n\rightarrow \infty} \mathbb{E}_\texttt{G} [ h_\texttt{G}^{\textrm{BP},\ell}(\varepsilon)]\, .
\end{equation}
The curve is zero until $\varepsilon = \varepsilon^\text{BP}$, at which point it jumps to a non-zero value and continues smoothly until it reaches one at $\varepsilon = 1$.
An upper bound for the MAP threshold $\bar{\varepsilon}^{\textrm{MAP}}$ can then be obtained by searching the unique value in 
$[\varepsilon^\text{BP},1]$ such that $\int_{\bar{\varepsilon}^{\text{MAP}}}^1 h^{\textrm{BP}}(\varepsilon)  \text{d}\varepsilon=r$. Operationally, we integrate the
curve $h^{\textrm{BP}}(\varepsilon)$ starting at $\varepsilon=1$ until the area under the curve is equal to the design rate of the code.
Since we consider regular ensembles whose BP EXIT curves jump at most once, this bound is conjectured to be tight~\cite{MeMoUr08}. (For general ensembles, a tighter bound can be achieved by using the extended BP EXIT curve~\cite{MeMoUr08}.)

\subsection{Density evolution of nonbinary LDPC codes}
\label{s:AsymptoticAnalysis}


The messages exchanged in the BP decoding are real vectors of length $2^m$, the $i$th element of which represents the a posteriori probability that the symbol is $i$.
In~\cite{RaUr05}, it was shown that in the case of transmission over the BEC the performance does not depend on the transmitted codeword 
and hence without loss of generality the transmission of the all-zero codeword can be considered. Under this assumption, 
the messages arising in the BP decoder assume a simplified form. In particular, the non-zero entries of a message are all equal and the message itself 
is equivalent to a subspace of  $\textrm{GF}_2^m$. 
The number of different subspaces of dimension $k$ of $\textrm{GF}_2^m$ is given by the Gaussian binomial coefficient,
\begin{equation}\label{e:gauss_bin_coeff}
{\GaussBin}_{m,k}=\left[\!\!\! \begin{array}{c} m\\ k \end{array} \!\!\!\right]= 
\begin{cases}
1 & \text{if } k=m \text{ or } k=0,\\
\displaystyle \prod_{\ell=0}^{k-1}\frac{2^m-2^\ell}{2^k-2^\ell} & \text{otherwise. } 
\end{cases}
\end{equation}
Since the non-zero elements of a message are equal, it is sufficient to keep track of the 
dimension of the messages~\cite{RaUr05}. 
We say that a message has dimension $k$ if it has $2^k$ non-zero elements. If a message coming from a node 
has dimension $k$, it means that the symbol is known to be one out of $2^k$ possible symbols or, equivalently, that at that node $m-k$ 
relations on the bits composing the symbol are known. Let us consider the three subspaces of dimension one of $\textrm{GF}_2^2$, $S_1=\{00,01\}$, $S_2=\{00,10\}$ and $S_3=\{00,11\}$. 
Subspaces $S_1$ and $S_2$ are representative of the case where one bit has been recovered and the other is still erased, while $S_3$ represents the case 
where the two bits are erased but their sum modulo-2 is known.

Let $P^{(\ell)}_\texttt{c}(k,d_\texttt{c})$ be the probability that a randomly chosen message computed by a check node and directed to a connected 
variable node at the 
$\ell$th iteration has dimension $k$, and let $P^{(\ell)}_\texttt{v}(k,d_\texttt{v})$ be the probability that a randomly chosen message computed by a variable 
node and directed to a check node at the $\ell$th iteration has dimension $k$.
At the check nodes, the BP decoder computes the sum of the subspaces corresponding to the incoming messages. We have the 
following recursion for $d_\texttt{c}\geqslant 3$ and $c=4,\cdots, d_\texttt{c}$~\cite{RaUr05}
\begin{equation}\nonumber
 P^{(\ell)}_\texttt{c}(k,3)=\sum_{i=0}^k \sum_{j=k-i}^k C^m_{i,j,k} P_\texttt{v}^{(\ell)} (i,d_\texttt{v}) P_\texttt{v}^{(\ell)}(j,d_\texttt{v})
\end{equation}
\begin{equation}\nonumber
 P^{(\ell)}_\texttt{c}(k,c)=\sum_{i=0}^k\sum_{j=k-i}^k C^m_{i,j,k}   P_\texttt{c}^{(\ell)} (i,c-1)P_\texttt{v}^{(\ell)}(j,d_\texttt{v})\, ,
\end{equation}
where $C^m_{i,j,k}=\frac{{\GaussBin}_{m-i,m-k} {\GaussBin}_{i,k-j} 
2^{(k-i)(k-j)}}{{\GaussBin}_{m,m-j}}$ is the probability of choosing a subspace of dimension $j$ whose sum with a subspace of dimension $i$ has dimension $k$.

At variable nodes, the decoder computes the intersection of the subspaces corresponding to the incoming messages. We denote by $P_\varepsilon(i)$ the probability 
that the message coming from the channel has dimension $i$, which is equivalent to the probability that $i$ bits are erased by the channel, and we have
$$
P_\varepsilon(i) =\binom{m}{i} \varepsilon^i (1-\varepsilon)^{m-i} , \; i=0,\cdots, m\, .
$$
We have the following recursion for $d_\texttt{v}\geqslant 2$ and $v=3,\cdots, d_\texttt{v}$~\cite{RaUr05}
\begin{equation}\nonumber
 P^{(\ell+1)}_\texttt{v}(k,2)=\sum_{i=k}^m  \sum_{j=k}^{m-i+k} V^m_{i,j,k}P_\varepsilon(i) P_\texttt{c}^{(\ell)}(j,d_\texttt{c})
\end{equation}
\begin{equation}\nonumber
 P^{(\ell+1)}_\texttt{v}(k,v)=\sum_{i=k}^m \sum_{j=k}^{m-i+k} V^m_{i,j,k} P^{(\ell+1)}_\texttt{v}(i,v-1) P_\texttt{c}^{(\ell)}(j,d_\texttt{c})\, ,
\end{equation}
where $V^m_{i,j,k}=\frac{{\GaussBin}_{i,k} {\GaussBin}_{m-i,j-k} 
2^{(i-k)(j-k)}}{{\GaussBin}_{m,j}}$ is the probability of choosing a subspace of dimension $j$ whose intersection with a 
subspace of dimension $i$ has dimension $k$. 

The asymptotic BP threshold is the largest channel parameter such that the decoding is successful and can be found as
$$
\varepsilon^{\text{BP}}=\text{sup}\{ \varepsilon \in[0,1]: P^{(\ell)}_\texttt{v}(0,d_\texttt{v})\xrightarrow{ \ell\rightarrow \infty}1\}\,.
$$
\section{BP-EXIT curve and MAP threshold}
\label{s:MAP threshold}

%
To draw the asymptotic average BP EXIT curve from~(\ref{e:BP EXIT}) we need to compute the bit erasure probabilities of the extrinsic BP decoder $P(\hat{x}^\ell_i=?| \yv_{\sim i})$.
In this section, starting from the DE equations in the previous section, we obtain the expression of the extrinsic messages delivered by 
the BP decoder. We then propose a method to compute the extrinsic bit probability $P(\hat{x}^\ell_i=?| \yv_{\sim i})$ to draw the 
BP EXIT curve for arbitrary $m$.

We can obtain the extrinsic symbol estimate of the BP decoder $\Psi_\texttt{ext}^{(\ell)}$ at the \mbox{$\ell$th} iteration 
taking into account all incoming messages to a variable node from the connected check nodes. Notice that due to the extrinsic nature of the message, the channel observations do not contribute in its computation.
We define $P^{(\ell)}_\texttt{ext}(k)$, $k=0,\cdots,m$, the probability that the message $\Psi_\texttt{ext}^{(\ell)}$ has dimension 
$k$ and we have the following recursion for $d_\texttt{v}\geqslant 2$ and $v=3,\cdots, d_\texttt{v}$
 \begin{equation}\nonumber
  P^{(\ell+1)}_\texttt{ext}(k,2)=\sum_{i=k}^m  \sum_{j=k}^{m-i+k} V^m_{i,j,k}P_\texttt{c}^{(\ell)}(i,d_\texttt{c}) P_\texttt{c}^{(\ell)}(j,d_\texttt{c})
 \end{equation}
 \begin{equation}\nonumber
  P^{(\ell+1)}_\texttt{ext}(k,v)=\sum_{i=k}^m \sum_{j=k}^{m-i+k} V^m_{i,j,k} P^{(\ell+1)}_\texttt{ext}(i,v-1) P_\texttt{c}^{(\ell)}(j,d_\texttt{c})\,.
 \end{equation}
Finally, we have 
 \begin{equation}\nonumber
  P^{(\ell+1)}_\texttt{ext}(k)=P^{(\ell+1)}_\texttt{ext}(k,d_\texttt{v})\,.
 \end{equation}
To compute~(\ref{e:BP EXIT}), we let the number of decoder iterations go to infinity. We define the following asymptotic quantities
$$
P_\texttt{ext}(k)=\lim_{\ell\rightarrow\infty}P^{(\ell)}_\texttt{ext}(k)
$$
$$
\hat{x}_i=\lim_{\ell\rightarrow\infty}\hat{x}^{(\ell)}_i\, ,\quad \Psi_\texttt{ext}=\lim_{\ell\rightarrow\infty}\Psi^{(\ell)}_\texttt{ext}
$$
and obtain
 \begin{equation}\label{e:Pb}
  P(\hat{x}_i=?| \yv_{\sim i})=\sum_{k=0}^m P(\hat{x}_i=?|\textrm{dim}(\Psi_\texttt{ext})=k, \yv_{\sim i})P_\texttt{ext}(k)\, .
 \end{equation}
Thanks to the tree assumptions, the probabilities in~(\ref{e:Pb}) do not depend on $i$ and~(\ref{e:Pb}) is actually the BP extrinsic entropy of a 
bit $h^{\textrm{BP}}(\varepsilon)$. To evaluate the probabilities \mbox{$P(\hat{x}_i=?|\textrm{dim}(\Psi_\texttt{ext})=k, \yv_{\sim i})$}, 
we have to enumerate the subspaces associated to the message $\Psi_\texttt{ext}$, so that we can compute 
 \begin{equation}\nonumber
  P(\hat{x}_i=?|\textrm{dim}(\Psi_\texttt{ext})\!=\!k, \yv_{\sim i})\!=\!\frac{1}{\left[\!\!\! \begin{array}{c} m\\ k \end{array} \!\!\!\right]}
\!\sum_{z=1}^{\scriptsize\left[\!\!\! \begin{array}{c} m\\ k \end{array} \!\!\!\right]} P^{(k)}_z(\hat{x}_i=?|\yv_{\sim i})\, ,
 \end{equation}
where $P^{(k)}_z(\hat{x}_i=?|\yv_{\sim i})$ is the extrinsic bit erasure probability associated to the $z$th subspace of dimension $k$, given an arbitrary 
but fixed ordering.

We propose an efficient method to identify the subspaces and to derive the corresponding extrinsic erasure probability. Note that each subspace of dimension
$k$ can be interpreted as the set of $2^k$ codewords of an $(m,k)$ binary linear block code 
of length $m$. Therefore, we can associate to each subspace of dimension $k$ an $(m-k)\times m$ matrix (the parity-check matrix of the code) such that the symbols corresponding to non-zero entries in the message 
belong to the nullspace of the matrix, meaning that the subspace associated to the message is the nullspace of the matrix. Thus, finding all subspaces of a certain dimension $k$ reduces to find all parity-check matrices that generate a different $(m,k)$ code. An efficient way to find these matrices is to find all $(m-k)\times m$ binary matrices in row-reduced echelon form containing no zero rows. The set of these matrices is denoted here as $\mathcal{R}_{m,k}$. 
A row-reduced echelon binary matrix is defined as a matrix in which (i) the first one in every 
row is in a column where all other elements are zero and (ii) the number of leading zeros increases in every row.
The nature of these matrices ensures that their nullspaces are distinct. Thus, the number of matrices in $\mathcal{R}_{m,k}$ corresponds 
to the number of different subspaces (or codes) of dimension $k$ of $\textrm{GF}_2^m$.
These matrices can be efficiently found by using a modified version of the algorithm described in~\cite{KnAg96}. 
\example
There are seven different subspaces of dimension one of $\textrm{GF}_2^3$, and we associate to them the matrices in $\mathcal{R}_{3,1}$
 \begin{eqnarray}\nonumber
\mathcal{R}_{3,1}&=\Bigg\{ \left[ \!\!  \begin{array}{c c c} 1 \!\!\!\!& 0 \!\!\!\!& 0\\ 0 \!\!\!\!& 1 \!\!\!\!& 0  \end{array} \!\!\right], 
 \left[ \!\!  \begin{array}{c c c} 1 \!\!\!\!& 0 \!\!\!\!& 0\\ 0 \!\!\!\!& 1 \!\!\!\!& 1  \end{array} \!\!\right],
 \left[ \!\!  \begin{array}{c c c} 1 \!\!\!\!& 0 \!\!\!\!& 0\\ 0 \!\!\!\!& 0 \!\!\!\!& 1  \end{array} \!\!\right], 
 \left[ \!\!  \begin{array}{c c c} 0 \!\!\!\!& 1 \!\!\!\!& 0\\ 0 \!\!\!\!& 0 \!\!\!\!& 1  \end{array} \!\!\right],\\ \nonumber
& \left[ \!\!  \begin{array}{c c c} 1 \!\!\!\!& 0 \!\!\!\!& 1\\ 0 \!\!\!\!& 1 \!\!\!\!& 1  \end{array} \!\!\right], 
 \left[ \!\!  \begin{array}{c c c} 1 \!\!\!\!& 0 \!\!\!\!& 1\\ 0 \!\!\!\!& 1 \!\!\!\!& 0  \end{array} \!\!\right],
 \left[ \!\!  \begin{array}{c c c} 1 \!\!\!\!& 1 \!\!\!\!& 0\\ 0 \!\!\!\!& 0 \!\!\!\!& 1  \end{array} \!\!\right]\Bigg\}.
 \end{eqnarray}

The next step consists of the computation of the erasure probability associated to each subspace. 
We know that \mbox{$P(\hat{x}_i=?|\textrm{dim}(\Psi_\texttt{ext})=0, \yv_{\sim i})=0$} since if the subspace has dimension 0 
the symbol is perfectly recovered after BP decoding. Also, \mbox{$P(\hat{x}_i=?|\textrm{dim}(\Psi_\texttt{ext})=m, \yv_{\sim i})=1$} 
since in this case we have complete uncertainty on the transmitted symbol.
The cases $k=1,\cdots, m-1$ are less intuitive. However, this problem is easy to solve by interpreting each subspace as an $(m,k)$ code with corresponding parity-check matrix in $\mathcal{R}_{m,k}$ and computing the probability of 
erasure assuming transmission over the BEC with erasure probability $\varepsilon$.
Since the length of these codes is generally very short, we can evaluate the erasure probability by using the full complexity decoding algorithm. 
Given the $z$th subspace of dimension $k$, whose nullspace is generated by the matrix $R^z$, 
we compute $P^{(k)}_z(\hat{x}_i=?|\yv_{\sim i})$ as~\cite{RiUr08}
\begin{equation}\nonumber
 \sum_{\mathcal{E}\subseteq [m]\smallsetminus\{i\}}\varepsilon^ {|\mathcal{E}|}(1-\varepsilon)^{m-1-|\mathcal{E}|}
(1+\textrm{rank}(R^z_\mathcal{E})-\textrm{rank}(R^z_{\mathcal{E}\cup i})),
\end{equation}
where $\mathcal{E}\subseteq [m]=\{1,\cdots m\}$ denotes the index set of erasures and $R^z_\mathcal{E}$ denotes the submatrix of $R^z$ 
indexed by the elements of $\mathcal{E}$.
In Table~\ref{t:h BP}, we report the expression of $h^{\textrm{BP}}(\varepsilon)$ for $m=1,\cdots, 4$. 
\begin{table*}[!t]
  \setlength{\tabcolsep}{4.0pt}
  \scriptsize \centering \caption{Expression of $h^{\textrm{BP}}(\varepsilon)$ for $m=1,\cdots,4$.  $G_{m,k}$ is the Gaussian binomial coefficient.}
  \label{t:h BP}
  \def\Hline{\noalign{\hrule height 2\arrayrulewidth}}
   \vskip -3.0ex 
 \begin{center} \begin{tabular}{clc}
  \Hline \\ [-2.0ex]
      $m$ & $h^{\textrm{BP}}(\varepsilon)$&\\
  \hline
  \\ [-2.0ex] \hline  \\ [-2.0ex]
      $1$  & $P_\texttt{ext}(1)$&\\
      $2$  & $\left( 1+\varepsilon\right)\frac{P_\texttt{ext}(1)}{G_{1,2}}+P_\texttt{ext}(2)$&\\
      $3$  & $\left(1+2\varepsilon+\varepsilon^2\right)\frac{P_\texttt{ext}(1)}{G_{3,1}}+\left(3+4\varepsilon-\varepsilon^2\right)\frac{P_\texttt{ext}(2)}{G_{3,2}}+P_\texttt{ext}(3)$&\\
      $4$  & $\left(1+3\varepsilon+3\varepsilon^2+\varepsilon^3\right)\frac{P_\texttt{ext}(1)}{G_{4,1}}+\left(7+18\varepsilon+9\varepsilon^2-6\varepsilon^3\right)\frac{P_\texttt{ext}(2)}{G_{4,2}}+
\left(7+12\varepsilon-6\varepsilon^2+\varepsilon^3\right)\frac{P_\texttt{ext}(3)}{G_{4,3}}$&\\\hline
  \end{tabular}\end{center}
  \vspace{-0.4cm}
  \end{table*}

\subsection{Results}
\begin{figure}
  \centering{}
  \includegraphics[width=8.8cm]{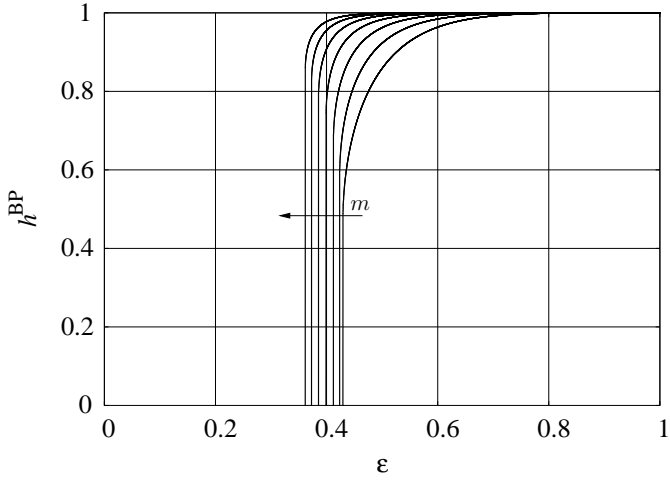} 
  \vspace{-0.7cm}
  \caption{BP EXIT curves for regular ensembles $\mathcal{G}(3,6,m)$, with $m=1,\cdots,7$.\label{f:BP36}}
  \vspace{-0.5cm}
  \end{figure}

In Fig.~\ref{f:BP36}, we plot the asymptotic average BP EXIT curves for the regular ensembles $\mathcal{G}(3,6,m)$, with $m=1,\cdots, 7$. 
We recall that the BP threshold $\varepsilon_\text{BP}$ is the point at which the curve jumps to a non-zero value. The figure shows that the BP threshold 
decreases as the alphabet size increases. This is in line with previous results in the literature, which state that the performance of the BP decoder 
degrades by moving to nonbinary alphabets when the variable node degree distribution is greater than two~\cite{RaUr05}. 

On the other hand, the MAP performance of the considered ensembles seems to improve when $m$ increases. In Table~\ref{t:Threshold}, we list the 
BP thresholds and the upper bounds on the MAP threshold $\bar{\varepsilon}_\text{MAP}$. 
Note that the $\bar{\varepsilon}_\text{MAP}$ rapidly increases with $m$ and approaches the channel capacity. Similar results were obtained for other ensembles.
 \begin{table}[!t]
 \setlength{\tabcolsep}{4.0pt}
 \scriptsize \centering \caption{BP and MAP thresholds for ensembles $\mathcal{G}(3,6,m)$. }
 \label{t:Threshold}
 \def\Hline{\noalign{\hrule height 2\arrayrulewidth}}
 \begin{tabular}{lccc}
 \Hline \\ [-2.0ex]
   $m$     & $\varepsilon_\text{BP}$ & $\bar{\varepsilon}_\text{MAP}$&  \\
 \hline
 \\ [-2.0ex] \hline  \\ [-2.0ex]
  1 & 0.42944  &  0.48815   & \\
  2 & 0.42347  &  0.49487   &\\
  3 & 0.41220  &  0.49791   & \\
  4 & 0.39890  &  0.49920  & \\
  5 & 0.38547  &  0.49970    &\\
  6 & 0.37288  &  0.499895  &\\
  7 & 0.36154  &  0.499965  &\\
  \hline
 \end{tabular}
 \vspace{-0.4cm}
 \end{table}

\section{Nonbinary Spatially-coupled LDPC Codes}
\begin{figure}
  \centering{}
  \includegraphics[width=0.7\columnwidth]{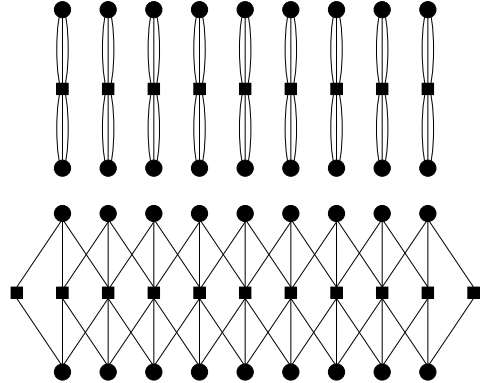} 
  \vspace{-0.2cm}
  \caption{Chains of nine protographs for the regular ensemble $\mathcal{G}(3,6,m)$. Top: non-interacting protographs. 
Bottom: a coupled chain of protographs.\label{f:protograph}}
\vspace{-0.5cm}
  \end{figure}

We consider nonbinary SC-LDPC code ensembles similar to the ensembles defined in~\cite{LeFeZiCo09} for binary codes, which are derived from regular convolutional protographs by termination. 
Consider as an example the coupling of regular codes with $d_\texttt{v}=3$ and $d_\texttt{c}=6$. 
The protograph for a $\mathcal{G}(3,6,m)$ ensemble 
is composed of two variable nodes and one check node, as shown in Fig.~\ref{f:protograph}, at the top. 
We consider a chain of $L$ protographs, adding a \textit{spatial} dimension to our code. Since these graphs do not interact, 
the chain behaves like the original $(3,6)$ regular code. An SC-LDPC code is then obtained by coupling the protographs: We connect each protograph to one neighboring protograph on the left, 
and to one neighboring protograph on the right, as shown in Fig.~\ref{f:protograph}, at the bottom. 
We denote this coupled ensemble as $\mathcal{G}_C(3,6,m,L)$.
Locally, the connectivity does not change with respect to the underlying ensemble. The only difference is at the boundaries, where one check 
node is added on each side to terminate the chain. 
Check nodes at positions $i\in[1,L-2]$ have degree six, while the degree of the remaining check nodes decreases linearly according to their position.
Having lower degree check nodes helps the decoder, at the expense of a loss in terms of design rate, which is reduced to $ r_{3,6}(L)=\frac{1}{2}-\frac{1}{L}$~\cite{KuRiUr11}.  
By increasing $L$, the rate loss is reduced, while the beneficial effect of low-degree check nodes on the BP performance does not vanish.

To draw the BP EXIT curve for the coupled ensemble, we apply the DE described in Section~\ref{s:AsymptoticAnalysis} to each section $i\in[0,L-1]$, taking 
into account of the \textit{spatial} structure of the code.
We then compute the extrinsic symbol estimate for each section and the corresponding BP extrinsic bit entropy. The BP EXIT curve of the SC ensemble is finally 
obtained by averaging over the $L$ entropies of the chain.

\begin{figure}
  \centering{}
  \includegraphics[width=8.8cm]{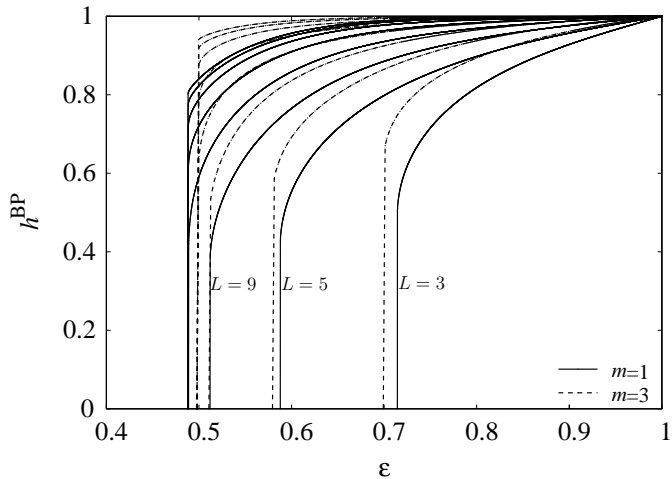} 
\vspace{-0.7cm}
  \caption{BP EXIT curves for SC ensembles $\mathcal{G}_C(3,6,m,L)$. Comparison between binary ($m=1$) and nonbinary ($m=3$) codes, for 
\mbox{$L=3,5,9,17,33,65,129$ and $257$}. \label{f:SC12}}
\vspace{-0.3cm}
  \end{figure}

In Fig.~\ref{f:SC12}, we show the BP EXIT curves of the ensembles $\mathcal{G}_C(3,6,m,L)$ for $m=1$ and $m=3$, for several values of~$L$. The corresponding 
BP and MAP thresholds for $m=3$ are given in Table~\ref{t:Threshold SC}, where we also report the Shannon limit $\varepsilon_\text{Sh}=1-r_{3,6}(L)$.
These results show that the same \textit{threshold saturation} effect observed for binary SC-LDPC codes occurs for the nonbinary codes. In fact, the BP threshold 
for large $L$ approaches the MAP threshold of the regular ensemble $\mathcal{G}(3,6,3)$. 
Furthermore, since the MAP performance of the underlying ensemble for $m=3$ outperforms that of the binary one (see Table~\ref{t:Threshold}), the BP thresholds of the nonbinary SC-LDPC code \textit{saturate} to a better value. 

\begin{figure}
  \centering{}
  \includegraphics[width=8.8cm]{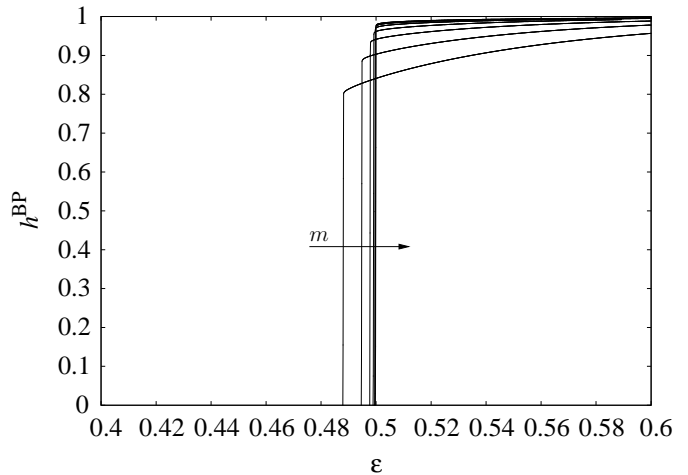} 
  \vspace{-0.7cm}
  \caption{BP EXIT curves for SC ensembles $\mathcal{G}_C(3,6,m,257)$, for $m=1,\cdots,7$. \label{f:SC1234}}
  \vspace{-0.5cm}
  \end{figure}

In Fig.~\ref{f:SC1234}, we report the BP EXIT curves of the ensembles $\mathcal{G}_C(3,6,m,257)$, for $m=1,\cdots,7$. 
The figure shows the interesting result that, contrary to non-coupled ensembles, the performance of SC-LDPC codes under BP decoding improves with $m$ and approaches the Shannon limit. Although not reported here due to lack of space, we obtained similar results for other ensembles.
\begin{table}[!t]
 \setlength{\tabcolsep}{4.0pt}
 \scriptsize \centering \caption{Asymptotic thresholds for ensembles $\mathcal{G}_C(3,6,3,L)$ in Fig.~\ref{f:SC12}. }
 \label{t:Threshold SC}
 \def\Hline{\noalign{\hrule height 2\arrayrulewidth}}
 \begin{tabular}{lcccc}
 \Hline \\ [-2.0ex]
   $L$     & $\varepsilon_\text{BP}$ & $\bar{\varepsilon}_\text{MAP}$& $\varepsilon_\text{Sh}$ &\\
 \hline
   \\ [-2.0ex] \hline  \\ [-2.0ex]
  3   & 0.69913  & 0.82738    & 0.83333&\\
  5   & 0.57947  & 0.68328    & 0.7   &\\
  9   & 0.51077  & 0.59026    & 0.61111&\\
  17  & 0.49795  & 0.54169    & 0.55882&\\
  33  & 0.49791  & 0.51847    & 0.53030&\\
  65  & 0.49791  & 0.50813    & 0.51538&\\
  129 & 0.49791  & 0.50272    & 0.50775&\\
  257 & 0.49791  & 0.50065    & 0.50389&\\
  \hline
 \end{tabular}
 \vspace{-0.4cm}
 \end{table}
\section{Conclusions}

We considered nonbinary SC-LDPC codes on the BEC. We proposed an efficient method to compute the BP EXIT curve for nonbinary LDPC codes and arbitrary alphabet size and subsequently an upper bound to the MAP threshold. Our analysis showed that performance under MAP decoding improves as the alphabet size increases. Furthermore, we analyzed nonbinary SC-LDPC code ensembles, showing that the same threshold saturation effect observed for binary SC-LDPC codes occurs
for nonbinary codes. Interestingly, the performance of nonbinary SC-LDPC codes under BP decoding improves with $m$ and approaches the Shannon limit, contrary to the case of nonbinary, non-coupled, LDPC codes. For instance, for the $\mathcal{G}_C(3,6,m,L)$ ensemble, the BP threshold is improved from 0.4881 for the binary case to 0.4997 for $m=5$ when $L$ tends to infinity.

Future work includes a formal proof of the threshold saturation phenomenon for nonbinary SC-LDPC codes, and a finite-length analysis.
  \section{Acknowledgements}
  The authors would like to thank Dr. Iryna Andriyanova for helpful discussions.  

\end{document}